\documentclass[lettersize,journal]{IEEEtran}

\usepackage{amsmath,amsfonts}
\usepackage{algorithmic}
\usepackage{algorithm}
\usepackage{array}
\usepackage{textcomp}
\usepackage{stfloats}
\usepackage{url}
\usepackage{verbatim}
\usepackage{graphicx}
\usepackage{xspace}
\usepackage[T1]{fontenc}

\usepackage{booktabs, tabularx, ragged2e}
\usepackage{multirow}
\usepackage{makecell}
\usepackage{xcolor}
\usepackage[most]{tcolorbox}
\usepackage{enumitem}
\usepackage[caption=false,font=normalsize,labelfont=sf,textfont=sf]{subfig}

\usepackage{cite} 

\usepackage{balance}

\makeatletter

\def\IEEEbibitemsep{0pt plus .5pt}
\let\old@openbib@code\@openbib@code
\renewcommand{\@openbib@code}{%
    \old@openbib@code
    \setlength{\itemsep}{0pt}%
    \setlength{\parsep}{0pt}%
    \setlength{\topsep}{0pt}%
    \setlength{\partopsep}{0pt}%
}
\makeatother

\begin{document}

\title{When Large Language Models Meet UAV Projects: An Empirical Study from Developers' Perspective}

\author{Yihua~Chen,
        Xingle~Que,
        Jiashuo~Zhang,
        Jiachi~Chen,
        Ting~Cui,
        Guangshun~Li,
        and~Ting~Chen
\thanks{Yihua Chen and Xingle Que contributed equally to this work.}
\thanks{Corresponding author: Ting Chen (e-mail: brokendragon@uestc.edu.cn).}
\thanks{Yihua Chen is with the School of Computer Science and Engineering (School of Cyber Security), University of Electronic Science and Technology of China, Chengdu 610054, China (e-mail: chenyihua@std.uestc.edu.cn).}%
\thanks{Xingle Que and Ting Chen are with the Shenzhen Institute for Advanced Study, University of Electronic Science and Technology of China, Shenzhen 518110, China (e-mail: 202522280609@std.uestc.edu.cn).}
\thanks{Jiashuo Zhang is with the School of Computer Science, Peking University, Beijing 100871, China.}
\thanks{Jiachi Chen is with The State Key Laboratory of Blockchain and Data Security, Zhejiang University, Hangzhou 310027, China.}
\thanks{Ting Cui is with the Institute of National Security and Development, Guangdong University of Finance and Economics, Guangzhou 510320, China (e-mail: cuiting@gdufe.edu.cn).}%
\thanks{Guangshun Li is with Qufu Normal University, Qufu 273165, China (e-mail: guangshunli@qfnu.edu.cn).}}



\maketitle

\bstctlcite{CTLsecret_diet}

\begin{abstract}
In recent years, unmanned aerial vehicles (UAVs) have become increasingly popular in our daily lives and have attracted significant research interest in software engineering. At the same time, large language models (LLMs) have made notable advancements in language understanding, reasoning, and generation, making LLM applications in UAVs a promising research direction. However, existing studies have largely remained in preliminary exploration with a limited understanding of real-world practice, which causes an academia-industry gap and hinders the application of LLMs in UAVs. 
This study aims to characterize common tasks and application scenarios of real-world UAV-LLM practices and investigate developers' perspectives to understand the challenges hindering the application of LLMs in the UAV domain.
We conducted the first large-scale empirical study involving 997 research papers and 1,509 GitHub projects. To gain deeper insights, we conducted a practitioner survey, receiving 52 valid responses from 15 countries to analyze differences between research efforts and industry practices.
The study classified nine common tasks (e.g., \textit{Natural Language Command Parsing}) in four UAV workflows (e.g., \textit{Information Input}) undertaken by LLMs in real-world UAV projects and revealed a large difference in the task distribution of research efforts and industry practices. While 40.4\% of developers have attempted to apply LLMs, 59.6\% still face integration challenges. These difficulties are attributed to five factors: technological maturity, performance, safety, cost, and others.
The findings highlight a clear gap between research and practice. This study provides practical implications for researchers and developers to better navigate technological and safety challenges in future UAV-LLM implementations.
\end{abstract}


\section{Introduction}
\label{sec:intro}

\IEEEPARstart{I}{n} recent years, unmanned aerial vehicles (UAVs) have become increasingly popular and play a vital role in various domains~\cite{shakhatreh2019unmanned,liu2021precision,colomina2014unmanned}.
This has attracted significant research interest in software engineering to implement advanced capabilities in UAV projects~\cite{wang2025dpfuzzer, khatiri2024simulation, minn2024dronlomaly, wang2025routhsearch}.
However, current UAV systems primarily rely on manual operation or rule-based pipelines~\cite{gupta2016survey}, limiting their flexibility and adaptability in dynamic environments. Consequently, achieving scalable high-autonomy operations is difficult, especially in complex scenarios requiring rapid reactions and varied tasks~\cite{colomina2014unmanned}.

Recent advances in large language models (LLMs) regarding reasoning and generation~\cite{chang2024survey,zhao2023llmsurvey} offer new possibilities to address these limitations. By enabling UAVs to incorporate reasoning and interpretation in complex environments, LLMs reduce reliance on manual pipelines~\cite{tian2025uavs,wang2024llmrobotics,zeng2023llmrobotics,ahn2022saycan,brohan2023rt2}, making UAV-LLM integration a prominent research direction. 

While existing studies~\cite{javaid2024large,tian2025uavs,tazir2023words} demonstrate feasibility, most remain confined to simulation or controlled environments, with limited coverage of scenarios and task diversity. They give insufficient consideration to real-world factors like computational resources, latency, and reliability~\cite{survey2023uavedge,wu2024safety}. This risks a mismatch between academic research and practical industrial needs.

To bridge this gap, this paper investigates three research questions (RQs):

\begin{description}
\item[\textbullet~RQ1.] \textit{What are the current research focuses and application scenarios of large language models in UAV systems?}

Identifying the specific tasks that LLMs can perform in UAV systems is the first step toward understanding the advantages and limitations of their integration. Through a systematic literature review (SLR) of 997 papers related to LLMs and UAVs using card sorting~\cite{spencer2009cardsorting}, we classify nine common tasks (e.g., \textit{Natural Language Command Parsing}) in four UAV workflows (e.g., \textit{Information Input}) undertaken by LLMs in real-world UAV projects. The most studied directions include \textit{Single-UAV Task Reasoning and Planning}, \textit{Control Logic and Command Generation}, and \textit{Natural Language Command Parsing}. This clarifies how LLMs are currently applied in UAV projects, enabling us to compare researchers' work with developers' practices in subsequent analysis.

\item[\textbullet~RQ2.] \textit{What are the differences between researchers' and developers' practices in integrating UAV projects with large language models?}

Researchers and developers often emphasize different aspects of new techniques depending on their objectives, resources, and evaluation standards. To explore these differences and better connect the two sides, we conducted an empirical analysis of 1,509 UAV-LLM-related projects on GitHub. Our analysis revealed divergent efforts: researchers focus on theoretical mechanisms and the integration of complex systems to improve autonomy, flexibility, and multitasking capabilities; in contrast, developers prioritize system stability, low-cost implementation, and rapid deployment through efficient solutions. This divergence, to some extent, limits the in-depth applications of LLMs in scenarios requiring high timeliness or resource-constrained operations. By revealing these practical differences, our results provide implications for facilitating real-world UAV-LLM practices.

\item[\textbullet~RQ3.] \textit{How do real-world developers perceive these differences, and what factors contribute to them?}

To clarify the root causes and solutions of the academia-industry gap, we designed a survey for real-world developers, focusing on the key obstacles and design choices they face when using LLMs for UAVs. The analysis of 52 real-world responses from 15 countries shows that the application of LLMs in UAVs remains at an early stage. Only 19.05\% of teams have deployed such projects, while 59.6\% have not attempted application, citing reasons such as \textit{existing technology has met the project requirements (54.8\%)}, \textit{insufficient performance of current LLMs (51.6\%)}, and \textit{security risks of large language models (45.2\%)}. The assessed technical maturity of all tasks is below three on a 5-point scale. Furthermore, 82.7\% of developers prefer \textit{Hybrid(Onboard Assistance) Modes} over direct control. These findings suggest that while researchers focus on expanding capabilities, developers are constrained by performance, reliability, and safety requirements, underscoring the need for closer alignment between research and practice.
\end{description}

In the following, we summarize the main contributions of this study as follows:
\begin{itemize}
    \item We identified and summarized nine types of tasks undertaken by LLMs in UAV projects and revealed the differences of their distributions in academic literature and industrial projects. The results reveal current applications and differences in LLM applications in UAVs among researchers and developers, providing a reference for future application-oriented research.
    
    \item Through a large-scale real-world empirical study, we analyzed developers' perspectives on LLM applications in UAVs. The findings identify potential reasons for the gap between researchers' and developers' practices and emphasize the need to bridge it.
    
    \item Our empirical results demonstrated challenges for future development and provided practical implications for researchers and developers. To support future research, our experimental data and analysis results are available at \url{https://github.com/UAVLLMs/OnlineSupplementMaterial}.
\end{itemize}

\section{Background}
\label{sec:background}

\subsection{Unmanned Aerial Vehicles}
Unmanned Aerial Vehicles (UAVs) are aerial platforms with perception, decision-making, and execution capabilities, widely used in various domains of our daily lives~\cite{tokekar2016sensor,ho2022open,erdemir2020autonomous,stolaroff2018energy,goodchild2018delivery}. Increasing mission complexity and dynamic environments drive a growing demand for UAVs for intelligence, including sophisticated control, efficient path planning, and natural human-computer interaction~\cite{shakhatreh2019unmanned}. To support this evolution, open-source autopilot platforms and communication frameworks have become an important foundation of UAV research and development. PX4~\cite{meier2015px4} provides a widely used flight control software stack that supports both research and industrial UAV applications. MAVLink~\cite{koubaa2019mavlink} provides a lightweight communication protocol for reliable message exchange between UAVs and ground stations. The ArduPilot project~\cite{baldi2022ardupilot} enables developers to build and customize UAV systems for diverse mission profiles.

\vspace{-1em}
\subsection{Large Language Models}

Large language models (LLMs) have demonstrated strong generalization and semantic understanding capabilities in natural language processing, code generation, automatic question answering, reasoning planning, and other tasks~\cite{naveed2023comprehensive}. Current mainstream LLMs (such as the ChatGPT~\cite{schulman2022chatgpt}, Claude~\cite{anthropic2024claude35}) are based on the Transformer architecture~\cite{abadi2015tensorflow}. They can extract deep semantics from natural language, perform logical reasoning, and model tasks. Furthermore, LLMs have acquired multimodal capabilities, showing broad prospects for integration with embedded systems in visual language understanding, robot control, and task planning.

\vspace{-1em}
\subsection{Card Sorting Method} 
\label{sec:card_sorting}

Card sorting is a cognitive organization method used to construct classification systems~\cite{wood2008card,spencer2009cardsorting}. In this process, researchers create cards for analysis objects (e.g., a document) that contain key information, such as functional descriptions or task instances. These cards are then classified bottom-up based on semantic associations to build a consistent classification system. Depending on whether categories are preset, the method is divided into three types: in \textit{closed card sorting}, all categories are preset and researchers only classify items into fixed categories; in \textit{open card sorting}, categories are not preset but are dynamically generated based on data characteristics during the sorting process; \textit{hybrid card sorting} combines the characteristics of both, allowing expansion based on existing categories to achieve both systematicity and flexibility.

\section{RQ1: What are the current research focuses and application scenarios of large language models in UAV systems?}
\label{sec:rq1}

LLMs offer powerful instruction understanding, task reasoning, and interactive control, promising enhanced intelligent perception and coordination for UAVs. However, existing studies often focus on specific models or cases, lacking a systematic classification of UAV-LLM integration. This gap limits the generalization of new techniques and hinders the exploration of cross-domain solutions for key problems like task planning and information interaction. Therefore, we conduct a systematic literature review (SLR) to identify current application fields, task categories, and future research directions.

\vspace{-1em}
\subsection{Systematic Literature Review (SLR)}
\label{sec:slr}

\begin{figure}[ht]
    \centering
    \includegraphics[width=250pt]{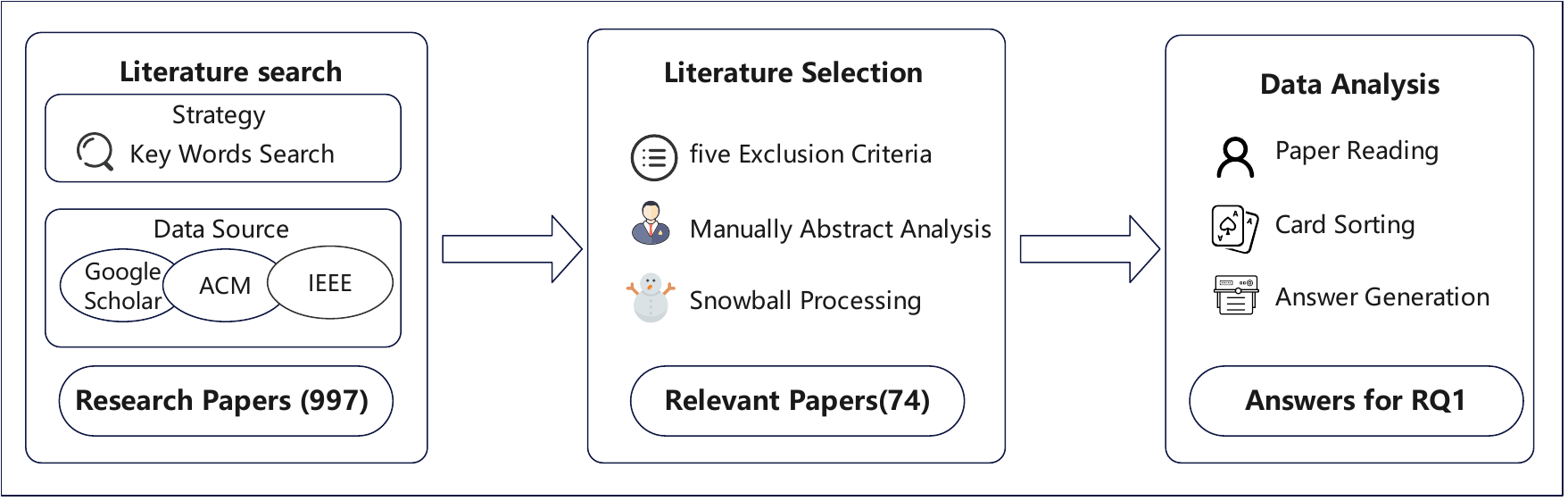}
    \caption{Methodological process for answering RQ1 via systematic literature review.}
    \label{fig:fig1}
\end{figure}

This study followed the systematic literature review (SLR) method guidelines proposed by Kitchenham ~\textit{et al.}~\cite{kitchenham2009systematic}. The specific process is shown in Figure~\ref{fig:fig1}, which includes three stages: \textit{literature search}, \textit{literature selection}, and \textit{data analysis}. First, we obtained 997 documents through keyword-based searches in Google Scholar, ACM, and IEEE. Secondly, following the five exclusion criteria, manual abstract analysis, and a snowball strategy~\cite{naderifar2017snowball}, we screened 74 documents relevant to the research question. Finally, through card sorting and inductive analysis, we categorized the selected literature to answer RQ1.

\subsubsection{Literature Search}
\label{sec:literature_search}

To collect high-quality research literature on UAV-LLM integration, we conducted a preliminary search across three widely-used digital libraries, namely \textit{Google Scholar}, \textit{ACM Digital Library}, and \textit{IEEE Xplore}, which together provide broad coverage of both software engineering and artificial intelligence venues. We executed identical title-level Boolean queries on each platform, requiring the paper title to contain at least one of the keywords ``Unmanned Aerial Vehicle (UAV)'' or ``Large Language Model (LLM)''. The retrieved records from the three sources were then merged, and duplicate entries (i.e., records sharing the same DOI or normalized title) were removed, yielding 997 unique candidate papers. To exclude literature that is not simultaneously related to both of these topics, such as UAV hardware platform design or photogrammetry applications~\cite{ahmed2022survey,shakhatreh2019unmanned,colomina2014unmanned}, we further required that their abstracts explicitly mention both ``UAV'' and ``LLM''.
\subsubsection{Literature Selection}
\label{sec:literature_selection}

To further filter out the literature that falls outside the scope of UAV-LLM integration. Papers that mention LLMs only as motivation without involving actual integration, application, or evaluation were excluded. We eliminated irrelevant articles based on five criteria: 

\begin{enumerate}[label=\textnormal{(\arabic*)}]
  \item Papers not written in English;
  \item Master's or doctoral dissertations;
  \item Research topics that are generalized and do not focus on the combination of UAVs and LLMs;
  \item Research that is not related to LLMs;
  \item Research that is not related to UAVs.
\end{enumerate}

To reduce bias in the filtering process, two graduate students independently assessed the relevance of each paper using a closed-card classification (``relevant'' vs.\ ``irrelevant''). Before any disagreement was resolved through discussion, we computed Cohen's Kappa coefficient~\cite{cohen1960coefficient} to measure the inter-rater reliability of the two annotators, obtaining $\kappa = 0.83$, which indicates almost-perfect agreement according to the benchmarks of Landis and Koch~\cite{landis1977measurement}. For the remaining inconsistent judgments, the two annotators reached a consensus through discussion, and a third senior researcher was consulted when consensus could not be reached. We further expanded the dataset via snowball sampling~\cite{naderifar2017snowball}. After multiple rounds of filtering, we selected 74 representative papers for further task categorization and quantitative analysis.

\subsubsection{Data Analysis}
\label{data_analysis}
Following literature filtering, we systematically analyzed the 74 selected papers using open-card classification to identify unknown task categories of LLMs in UAV systems. Two researchers independently read each paper and extracted LLM applications, initially identifying raw expressions (e.g., ``natural language parsing of flight commands'' and ``dialog-based multi-round path replanning'') and then condensing them into standardized task intents (e.g., ``task parsing'' and ``path reasoning and planning''). Prior to consolidating the two researchers' independent labels, we again computed Cohen's Kappa to evaluate the inter-rater reliability of the task-intent extraction, obtaining $\kappa = 0.79$, which corresponds to substantial agreement~\cite{landis1977measurement}. The remaining discrepancies were then resolved through multiple rounds of discussion. The standardized task intents were grouped into higher-level categories based on semantic relevance and technical logic, yielding nine task types that were abstracted into four core UAV workflow stages, ensuring consistency and reliability in the resulting classification system.

\vspace{-1em}
\subsection{Results}
\label{results}
\begin{figure}[h!]
\centering
\includegraphics[width=250pt]{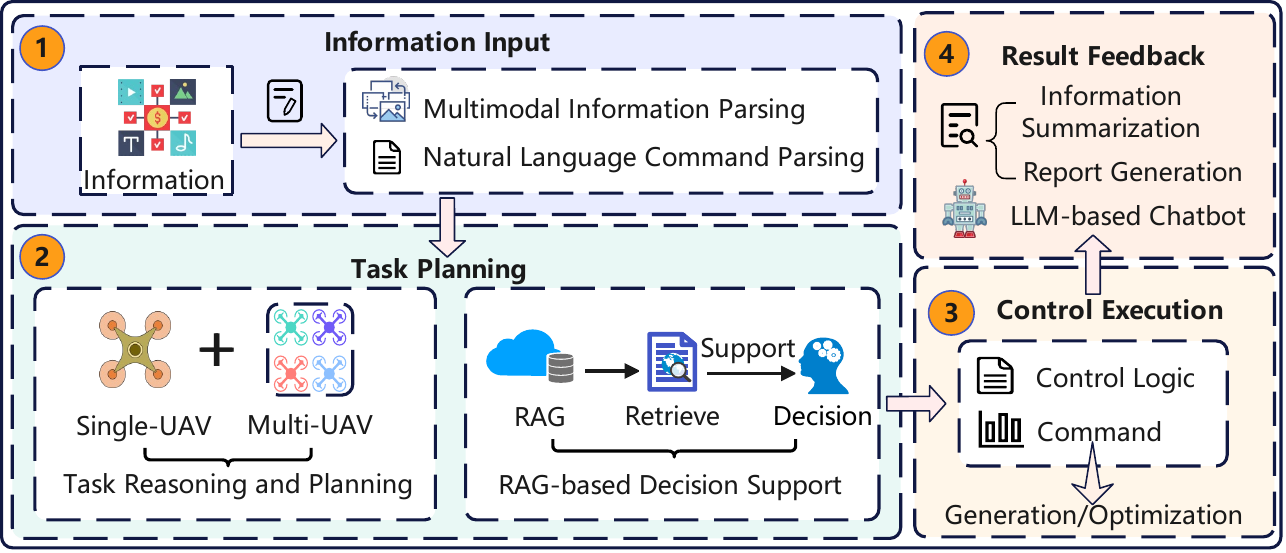}
\caption{Illustration of the UAV workflow and LLM task modules.}
\label{fig:fig3}
\end{figure}

Based on analysis results, we have summarized the UAV workflow into the following four steps: \textit{Information Input, Task Planning, Control Execution, and Result Feedback}, which are shown in Figure~\ref{fig:fig3}. Each step is associated with specific tasks that LLMs assist with. In the \textit{Information Input} phase, LLMs support \textit{Multimodal Information Parsing} and \textit{Natural Language Command Parsing}. In the \textit{Task Planning} phase, LLMs are involved in \textit{Single-UAV and Multi-UAV Task Reasoning and Planning}, as well as \textit{RAG-based Decision Support}. The \textit{Control Execution} phase utilizes \textit{Control Logic and Command Generation/Optimization}. Finally, the \textit{Result Feedback} phase incorporates \textit{Information Summarization and Report Generation} and the development of \textit{LLM-based Chatbot}.

Our analysis summarizes the nine tasks that LLMs currently undertake in UAV systems (see Table~\ref{tab:tasks}). These nine tasks cover the UAV workflow from information input to task planning, control execution, and result feedback. Specifically, they include: \textit{Natural Language Command Parsing (33.78\%), Multimodal Information Parsing (31.08\%), Single-UAV Task Reasoning and Planning (39.19\%), Multi-UAV Task Reasoning and Planning (14.86\%), RAG-based Decision Support (20.27\%), Control Logic and Command Generation (36.48\%), Control Logic and Command Optimization (12.16\%), Information Summarization and Report Generation (25.68\%), and LLM-based Chatbot (20.27\%)}. These task modules, as part of the intelligent UAV system workflow, can support autonomous perception and decision-making in complex environments. It should be noted that a single paper may involve multiple tasks, so the sum of the proportions exceeds 100\%.

{\setlength{\dbltextfloatsep}{6pt plus 1pt minus 2pt}
\begin{table*}[t]
    \centering
    \small
    \renewcommand{\arraystretch}{1.35}
    \setlength{\tabcolsep}{5pt}        
    
    \caption{Tasks Performed by LLMs in UAV Systems.}
    \label{tab:tasks}
    \newcolumntype{L}{>{\raggedright\arraybackslash}X}
    \begin{tabularx}{\textwidth}{@{} p{2.5cm} p{4.2cm} L L @{}}
        \toprule
        \textbf{UAV Workflow} & \textbf{Task Name} & \textbf{Description} & \textbf{Example Application} \\
        \midrule

        \multirow{2}{*}{\textbf{Information Input}} 
        & Natural Language Command Parsing 
        & Converts user natural-language commands into structured task intents. 
        & ``Please fly over the building complex and take photos'' $\rightarrow$ \texttt{\{fly, to: coords, action: photo\}} \\
        \cmidrule(l){2-4} 
        
        & Multimodal Information Parsing 
        & Analyses and annotates UAV imagery, video, and remote-sensing data. 
        & Image $\rightarrow$ textual description, bounding-box targets, environmental semantic labels. \\
        \midrule

        \multirow{3}{*}{\textbf{Task Planning}} 
        & Single-UAV Task Reasoning and Planning 
        & Performs multi-stage task decomposition and environmental path planning through reasoning. 
        & Detect weather-affected routes and reschedule task order accordingly. \\
        \cmidrule(l){2-4}
        
        & Multi-UAV Task Reasoning and Planning 
        & Allocates tasks among multiple UAVs, dynamically adjusts plans, and executes cooperative control. 
        & Assign tasks to five UAVs and dynamically switch targets. \\
        \cmidrule(l){2-4}
        
        & RAG-based Decision Support 
        & Retrieves regulations, maps, and domain knowledge to enhance decision-making. 
        & Determine whether an area is a no-fly zone / find the energy-optimal route. \\
        \midrule

        \multirow{2}{*}{\textbf{Control Execution}} 
        & Logic and Command Generation 
        & Outputs low-level control code (e.g., PX4, MAVLink) according to the mission plan. 
        & Automatically generate flight-control scripts and inject them into the controller. \\
        \cmidrule(l){2-4}
        
        & Control Logic and Command Optimization 
        & Senses the spectrum environment and network topology to optimise communication strategies and ad-hoc networking. 
        & Select interference-free channels; rebuild relay networks in disaster zones. \\
        \midrule

        \multirow{2}{*}{\textbf{Result Feedback}} 
        & Information Summarization and Report Generation 
        & Extracts key points from UAV status, logs, and sensor data to produce concise summaries or reports. 
        & Generate a daily disaster-monitoring report from the logs of multiple UAVs. \\
        \cmidrule(l){2-4}
        
        & LLM-based Chatbot 
        & Manages tasks, status queries, and user feedback through question-and-answer dialogue. 
        & ``Have you completed the target?'' $\rightarrow$ ``The task in Area 2 is complete.'' \\
        \bottomrule
    \end{tabularx}
\end{table*}

\begin{center}
\begin{tcolorbox}[colback=gray!10,
                  colframe=black,
                  width=\columnwidth,
                  arc=1mm, auto outer arc,
                  boxrule=0.5pt,
                 ]
\noindent
\emph{\textbf{Finding 1:}  Academic research spans four UAV workflow areas: Information Input, Task Planning, Control Execution, and Result Feedback. The most prevalent tasks are Single-UAV Task Reasoning and Planning (39.19\%), Logic and Command Generation (36.48\%), Multimodal Information Parsing (31.08\%), Natural Language Command Parsing (33.78\%). }
\end{tcolorbox}
\end{center}

\begin{center}
\begin{tcolorbox}[colback=gray!10,
                  colframe=black,
                  width=\columnwidth,
                  arc=1mm, auto outer arc,
                  boxrule=0.5pt,
                 ]
\noindent
\emph{\textbf{Finding 2:}  Single-UAV Task Reasoning and Planning is the most studied task (39.19\%), while Control Logic and Command Optimization is the least (12.16\%). This relatively small difference in data suggests that researchers' attention across tasks is relatively dispersed, with no single area receiving close attention. }
\end{tcolorbox}
\end{center}

\subsubsection{Information Input}
The \textit{Information Input} module initiates the UAV workflow by converting user commands and perception data into machine-executable task representations~\cite{colomina2014unmanned,survey2023uavedge}. In this module, LLMs are mainly used for \textit{Natural Language Command Parsing} and \textit{Multimodal Information Parsing}.

\textbf{Natural Language Command Parsing.}
\textit{Natural Language Command Parsing} maps free-form user instructions to structured task intents. Compared with traditional speech/rule-based pipelines~\cite{gupta2016survey}, LLM-based methods better capture context and implicit intent, improving flexibility and execution accuracy~\cite{ahn2022saycan}. For example, commands such as ``\textit{Fly over the west side of the bridge and take a photo}'' can be decomposed into action, target location, and mission type. Representative systems include Swarm-GPT~\cite{jiao2023swarm}, TypeFly~\cite{chen2023typefly}, and AutoHMA-LLM~\cite{10839354}.

\textbf{Multimodal Information Parsing.}
\textit{Multimodal Information Parsing} uses LLM-centered vision-language pipelines to interpret UAV images, videos, and remote-sensing streams for downstream planning. Unlike traditional single-modality approaches~\cite{zhu2021survey}, recent systems combine visual encoders (e.g., BLIP-2, SAM, Grounding DINO) with V-L modeling~\cite{li2023blip2,kirillov2023sam,liu2023groundingdino,zeng2023vlmsurvey}, improving scene understanding and semantic extraction in tasks such as inspection and security monitoring. Examples include AeroReformer~\cite{li2025aeroreformeraerialreferringtransformer}, CLIPSwarm~\cite{10801327}, and Hawkeye~\cite{zhao2024hawkeye}.

\subsubsection{Task Planning}
\textit{Task Planning} converts mission objectives into executable plans and adapts them to dynamic environments through reasoning, path planning, and allocation~\cite{shakhatreh2019unmanned}. In our taxonomy, this module includes \textit{Single-UAV Task Reasoning and Planning}, \textit{Multi-UAV Task Reasoning and Planning}, and \textit{RAG-based Decision Support}.

\textbf{Single-UAV Task Reasoning and Planning.}
This task uses LLMs to decompose multi-stage missions and generate adaptive paths for a single UAV. Compared with rule-based or classical planners~\cite{zhu2019uavpathplanning}, LLM-based approaches better handle changing constraints (e.g., weather, obstacles, and priorities) via online reasoning~\cite{huang2024lmplanners}. Representative systems include NavAgent~\cite{liu2024navagent}, RL-integrated methods~\cite{10650538}, and TAMP-style frameworks~\cite{sautenkov2025uavvlavisionlanguageactionlargescale}.

\textbf{Multi-UAV Task Reasoning and Planning.}
This task focuses on collaborative assignment and re-planning across multiple UAVs. Traditional schedulers based on fixed rules often underperform in dynamic settings~\cite{chung2018multiuavsurvey}. LLM-based methods improve coordination by dynamically reallocating tasks and routes according to mission status and environmental changes. Examples include AutoHMA-LLM~\cite{10839354}, Swarm-GPT~\cite{jiao2023swarm}, and FlockGPT~\cite{10765199}.

\textbf{RAG-based Decision Support.}
This task combines LLMs with external knowledge for planning-time decision support. Unlike static rule bases, RAG~\cite{lewis2020rag} enables retrieval of regulations, maps, and no-fly constraints to improve contextual validity and compliance. Applications include SAGIN management~\cite{69} and regulation-aware planning scenarios~\cite{liu2024navagent,javaid2024large}.

\subsubsection{Control Execution}
The \textit{Control Execution} module translates planning outputs into executable flight actions, including command generation and runtime strategy adjustment under dynamic conditions~\cite{shakhatreh2019unmanned}.

\textbf{Control Logic and Command Generation.}
This task uses LLMs to convert mission plans into low-level UAV commands (e.g., PX4/MAVLink). Compared with manual scripting or predefined command templates~\cite{koubaa2019mavlink,mavlinkdevguide}, LLM-based generation reduces coding burden and improves development efficiency while preserving execution intent. Representative examples include TypeFly~\cite{chen2023typefly} and PX4-oriented command-generation studies~\cite{tazir2023words,56}.

\textbf{Control Logic and Command Optimization.}
This task optimizes control strategies from runtime feedback and environmental signals. Traditional fixed-rule tuning is often brittle in dynamic scenarios~\cite{campion2018swarmsurvey}. LLM-assisted methods support adaptive communication/link optimization (e.g., channel selection, anti-jamming, and topology adaptation), improving swarm robustness and stability~\cite{wei2025llmwirelesssurvey,zhou2024llmicl}. Typical applications include SAGIN network scheduling and security-oriented resource coordination~\cite{69,15,30,50}.

\subsubsection{Result Feedback}
The \textit{Result Feedback} module delivers post-execution outputs to users, including status summaries, report generation, and interactive feedback, thereby improving transparency and communication efficiency in UAV operations~\cite{wang2024mhri,zhang2024llmhri}.

\textbf{Information Summarization and Report Generation.}
This task uses LLMs to summarize UAV status, logs, perception streams, and trajectories into concise operational reports. Compared with manual reporting or conventional analytics, LLM-based pipelines improve timeliness and adaptability to dynamic missions by automatically extracting progress, key events, and potential risks. Representative examples include public-attitude analysis systems~\cite{23} and Hawkeye~\cite{zhao2024hawkeye}, where LLMs support real-time interpretation of complex monitoring data.

\textbf{LLM-based Chatbot.}
This task enables natural-language interaction between users and UAV systems for mission queries, progress tracking, and explanatory feedback. Unlike fixed interfaces, LLM-driven multi-turn dialogue provides more flexible and personalized interaction~\cite{zhang2024llmhri}. Typical systems support status querying, explanation generation, and prompt-based interaction~\cite{zhao2024multibotgpt,aikins2024leviosa}; these capabilities can be further strengthened by integrating voice and multimodal front-ends~\cite{park2023voiceuav,choutri2022multilingual,oneata2021multimodal}.

\section{RQ2: What are the differences between researchers' and developers' practices in integrating UAV projects with large language models?}
\label{sec:rq2}

To assess the current level of developers' practices in integrating LLMs with UAVs and to identify the main challenges, we compiled a dataset of real-world projects that combine them. We then compared and analyzed the collected papers and projects, using the four core application domains and nine task categories defined earlier, in order to clarify the gaps between researchers' and developers' practices and to explore the reasons for those gaps.

\vspace{-1em}
\subsection{Data Collection and Processing}
\label{sec:data_coll}

We mined GitHub to identify real-world UAV-LLM integration projects via keyword-based retrieval. To reduce false negatives, we constructed two curated keyword sets in two stages. First, we built a seed vocabulary from: (i) keywords in the 74 SLR papers selected in Section~\ref{sec:slr}; (ii) official names of widely used general-purpose and multimodal LLMs from major AI labs; and (iii) common UAV platform/airframe terms used in open-source flight-control communities (e.g., PX4, ArduPilot). Second, we iteratively refined this vocabulary through exploratory GitHub queries and manual inspection of returned repositories, adding frequently occurring model names and UAV aliases found in titles and READMEs.

The final sets were:
\begin{itemize}[leftmargin=1.2em,itemsep=0pt,topsep=0pt,parsep=0pt,partopsep=0pt]
  \item \textbf{UAV terms}: ``drone'', ``UAV'', ``unmanned aerial vehicle'', ``quadcopter'', ``multirotor'', ``autopilot'', ``PX4'', ``ArduPilot'', ``MAVLink'', ``Pixhawk'', ``aerial robot''.
  \item \textbf{LLM terms}: ``GPT'', ``ChatGPT'', ``GPT-4'', ``GPT-4V'', ``large language model'', ``LLM'', ``foundation model'', ``LLaMA'', ``Llama-2'', ``Llama-3'', ``Vicuna'', ``Mistral'', ``Mixtral'', ``Gemini'', ``Claude'', ``PaLM'', ``Qwen'', ``Qwen-VL'', ``LLaVA'', ``MiniGPT-4'', ``BLIP-2'', ``CLIP'', ``transformer'', ``vision-language model'', ``VLM'', ``multimodal large language model'', ``MLLM''.
\end{itemize}

A repository was retained if at least one UAV term and one LLM term co-occurred in its title, README, or description. We acknowledge residual keyword-selection bias (e.g., projects using only generic labels such as ``chatbot'', or newly released models after the retrieval cut-off). To mitigate this, we performed snowball expansion by inspecting related repositories and citation/link references of initially retrieved projects and merged newly identified relevant repositories into the candidate pool.

After deduplication, two authors independently screened all candidates. Before disagreement resolution, we measured inter-rater reliability using Cohen's Kappa~\cite{cohen1960coefficient} and obtained $\kappa=0.81$, indicating almost-perfect agreement~\cite{landis1977measurement}. Disagreements were resolved through discussion, yielding 56 projects with complete engineering implementations or prototype demonstrations.

To examine the research--practice gap in LLM-enabled UAV integration, we compared 74 SLR publications (research perspective) with 56 GitHub projects (developer perspective). Each instance was annotated using the RQ1 task-classification framework and mapped to the four core domains and nine specific tasks in Figure~\ref{fig:fig2}. We then compared task-frequency distributions across the two ecosystems. Annotation and reconciliation were conducted iteratively through multiple researcher discussions to ensure rigor and consistency.

\begin{figure*}[ht]
    \centering
    \includegraphics[width=460pt]{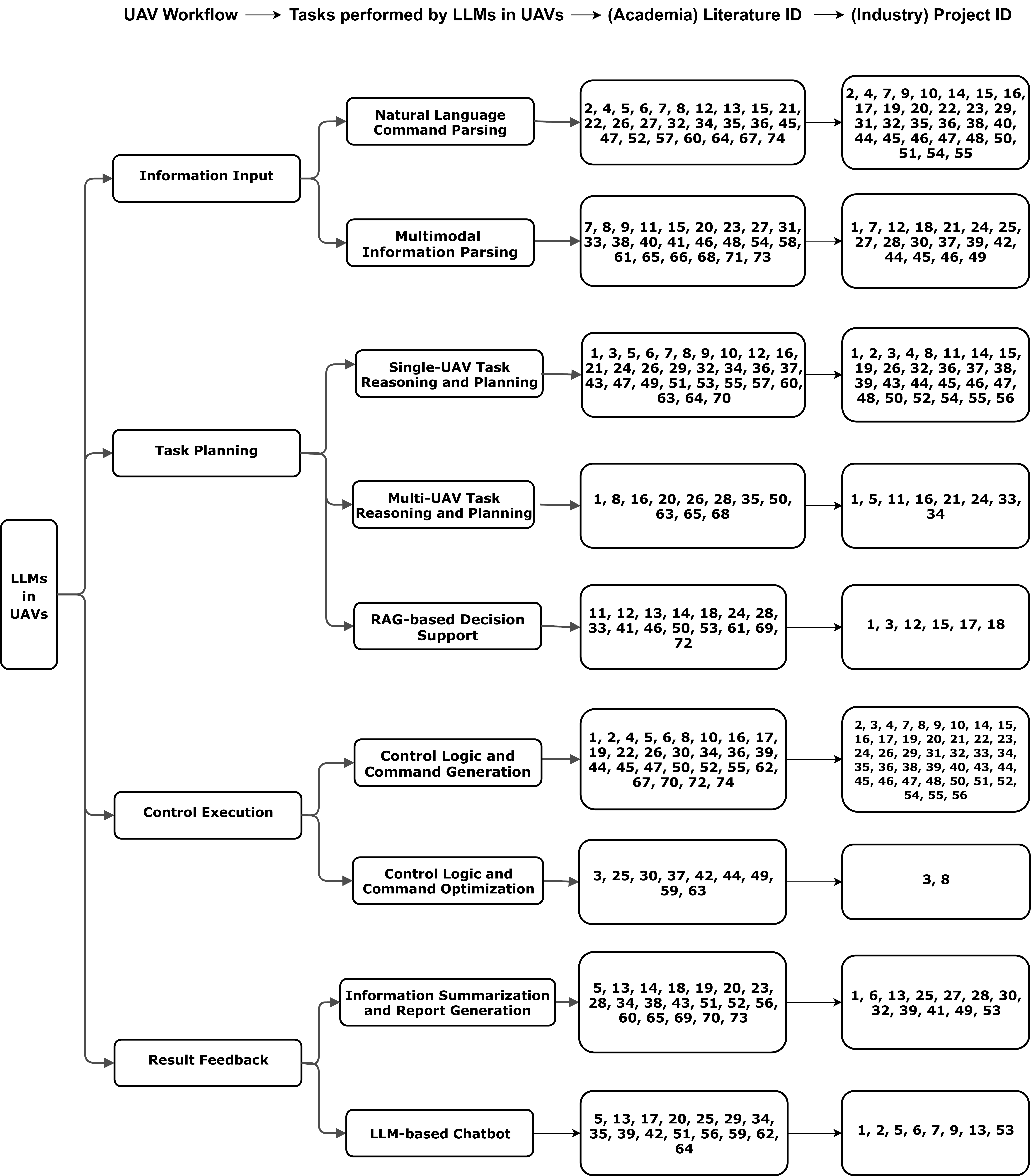}
    \caption{Statistics on LLM applications in UAVs between researchers and developers.}
    \label{fig:fig2}
\end{figure*}

\vspace{-1em}
\subsection{Results}
\subsubsection{Current Practices}
Firstly, we conducted a statistical analysis of the tasks performed by LLMs in UAV projects for each developer. The results include: \textit{Natural Language Command Parsing} (51.79\%), \textit{Multimodal Information Parsing} (30.36\%), \textit{single-UAV Task Reasoning and Planning} (50\%), \textit{Multi-UAV Task Reasoning and Planning} (14.29\%), \textit{RAG-based Decision Support} (8.92\%), \textit{Control Logic and Command Generation} (71.43\%), \textit{Control Logic and Command Optimization} (3.57\%), \textit{Information Summarization and Report Generation} (21.43\%), and \textit{LLM-based Chatbot} (14.29\%).
We observed that developers are most focused on flight control in UAV projects, with projects performing the task of \textit{Control Logic and Command Generation} accounting for 71.43\% of all projects. \textit{Natural Language Command Parsing} and \textit{Single-UAV Task Reasoning and Planning} account for 51.79\% and 50\%, respectively, indicating an increasing demand among developers for task automation and intelligent interaction. 
In contrast, \textit{RAG-based Decision Support} and \textit{Control Logic and Command Optimization} are less commonly applied, possibly because they require more external knowledge bases and datasets. Additionally, \textit{Multimodal Information Parsing} and \textit{Multi-UAV Task Reasoning and Planning} have seen some applications but still face challenges related to real-time performance, computational resource demands, and the inherent complexity of these tasks.

\begin{center}
\begin{tcolorbox}[colback=gray!10,
                  colframe=black,
                  width=\columnwidth,
                  arc=1mm, auto outer arc,
                  boxrule=0.5pt,
                 ]
\noindent
\emph{\textbf{Finding 3:} In over 50\% of the projects, the tasks handled by large models are concentrated in Control Logic and Command Generation (71.43\%), Natural Language Command Parsing (51.79\%), and single-UAV Task Reasoning and Planning (50\%). }
\end{tcolorbox}
\end{center}

\subsubsection{The differences between researchers and developers}
We analyzed the literature and projects that integrate UAVs with LLMs, presenting their specifics in Figure~\ref{fig:fig2} and identifying three key differences between researchers' and developers' practices in integrating UAV projects with LLMs. In Figure~\ref{fig:fig2}, ``UAV Workflow'' and ``Tasks performed by LLMs in UAVs'' represent the four UAV workflows and the nine tasks performed by LLMs in UAV projects that we identified in RQ1. ``(Researchers) Literature ID'' and ``(Developers) Project ID'' refer to the IDs of the literature and projects we collected. We have analyzed the UAV tasks involved in each of the literature and projects. 

\textbf{Prevalence and Distribution of Real-World UAV-LLM projects.}  In terms of prevalence, there are currently fewer developers' projects combining large language models with UAVs than academic research, which could be attributed to the field's early development stage and high development cost. Academic focus is dispersed across nine tasks (ranging from 12.16\% to 39.19\%), covering broad areas such as multi-UAV collaboration and decision support, without significant concentration. In contrast, developers' projects exhibit task centralization (ranging from 3.57\% to 71.43\%), with a focus on product interaction, practical automation, and execution.

\textbf{Application Fields.}  Researchers emphasize cutting-edge theoretical modeling and optimization, such as algorithm refinement, data fusion, and external knowledge utilization. For example, researchers address multi-UAV collaboration via RAG and reinforcement learning~\cite{liu2024navagent,javaid2024large}, optimize communication via GNNs and LLM reasoning~\cite{15}, and utilize LLMs for link-quality prediction~\cite{69}. Conversely, developers prioritize real-time requirements, operability, and scalability, focusing on improving task execution efficiency and control accuracy in real-world applications. For example, in more than 50\% of the projects, large models are responsible for tasks such as \textit{Natural Language Command Parsing}, \textit{Control Logic and Command Generation}, and \textit{Single-UAV Task Reasoning and Planning}. These tasks are concentrated in areas like flight control, task planning, and human-machine interaction.

\begin{center}
\begin{tcolorbox}[colback=gray!10,
                  colframe=black,
                  width=\columnwidth,
                  arc=1mm, auto outer arc,
                  boxrule=0.5pt,
                 ]
\noindent
\emph{\textbf{Finding 4:} Regarding the development of \textit{RAG-based Decision Support} and \textit{Control Logic and Command Optimization}, researchers are primarily focused on these areas, with respective proportions of 20.27\% and 12.16\%. In contrast, their application in developers' projects is relatively limited, with respective proportions of 8.92\% and 3.57\%.
}
\end{tcolorbox}
\end{center}

\textbf{Different UAV-LLM Integration Practices.}  To gain a deeper understanding of the technical details of how researchers and developers technically integrate LLMs into UAV projects, we identified three integration modes based on LLM location and interaction: \textit{External (Offboard Invocation) mode, Hybrid (Onboard Assistance) mode, and Internal (Onboard Control) mode}. We researched their distribution in current researchers' and developers' practices.

\begin{enumerate}[label=\textnormal{(\arabic*)}]
  \item \textbf{Offboard Invocation (External):} LLMs are deployed outside the UAV, either in the cloud or at a ground-based edge computing node. It receives task information from the UAV via remote APIs and returns task suggestions and semantic interpretations. This mode offers flexible deployment and strong computational power but suffers from high communication latency and limited real-time performance, making it better suited to non-real-time tasks such as large-scale inference and complex semantic parsing. The use of this mode by researchers (52.7\%) is higher than by developers (41.07\%), reflecting that researchers are largely focused on cloud-based invocation, with tasks concentrated on algorithm verification and offline task planning. In contrast, developers have reduced this proportion, migrating tasks to edge/onboard systems to enhance real-time performance.
  
  \item \textbf{Onboard Assistance (Hybrid):} LLMs are deployed on an edge computing device within the UAV (such as Jetson, RK3588, etc.). It does not directly generate control commands but acts as a decision-support module, collaborating with the UAV's flight control system to support task planning, path inference, and multi-UAV coordination. This mode balances intelligence with stability for real-time perception. The usage of this mode by developers (50\%) is higher than by researchers (38\%),  suggesting that real-world projects prioritize stability and real-time performance over extreme inference capabilities, while researchers are often limited by hardware constraints and deployment complexity.
  
  \item \textbf{Onboard Control (Internal):} LLMs are deeply integrated into the UAV's flight control system, directly responsible for task decomposition, action decision-making, and control command generation. Its advantage is that the UAV's flight control directly relies on the LLM, providing good real-time performance. It also offers high real-time capability and autonomy, making it a key architecture for achieving fully autonomous flight. However, the disadvantage is the high demand for local computational resources, which is limited by the model size. The usage of this mode is low among both researchers (5.4\%) and developers (10.71\%), reflecting a cautious attitude. This mode has significant growth potential as lightweight models and edge computing evolve.
  
\end{enumerate}

\begin{center}
\begin{tcolorbox}[colback=gray!10,
                  colframe=black,
                  width=\columnwidth,
                  arc=1mm, auto outer arc,
                  boxrule=0.5pt,
                 ]
\noindent
\emph{\textbf{Finding 5:} In the three integration methods of LLMs and UAVs, researchers most commonly adopt \textit{External-offboard invocation (52.7\%)}, while developers tend to prioritize \textit{Hybrid-onboard assistance (50\%)}. This indicates that researchers focus more on algorithm validation and performance evaluation in controlled environments, whereas developers place greater emphasis on real-time performance and the practical application of resource constraints.}
\end{tcolorbox}
\end{center}

\begin{center}
\begin{tcolorbox}[colback=gray!10,
                  colframe=black,
                  width=\columnwidth,
                  arc=1mm, auto outer arc,
                  boxrule=0.5pt,
                 ]
\noindent
\emph{\textbf{Finding 6:} Currently, LLMs are rarely directly involved in UAV control, with the usage of this mode being low among both researchers (5.4\%) and developers (10.71\%), reflecting a cautious attitude towards applying LLMs to core UAV functions. However, the development of edge computing and lightweight models may bring new possibilities for the application of large models in UAVs.}
\end{tcolorbox}
\end{center}

\section{RQ3: How do real-world developers perceive these differences, and what factors contribute to them?}

\label{sec:rq3}
Section 4 presents preliminary findings that reveal the differences in integrating UAVs with LLMs between researchers' and developers' practices. To further explore the potential reasons behind these differences and identify possible solutions to bridge the gap, we conducted an online survey targeting real-world UAV developers. The goal was to gain deeper insights into developers' perspectives on integrating LLMs into UAV systems, as well as to examine the key obstacles and design choices encountered in this integration process.

\vspace{-1em}
\subsection{Survey Design}
\label{sec:survey_design}
The UAV survey, designed following the guidelines for personal opinion surveys proposed by Kitchenham and Pfleeger~\cite{kitchenham2008personal}, adopts an anonymous format~\cite{tyagi1989effects} to encourage candid feedback and increase response rates. It is divided into four sections.

\noindent \textbf{Demographics.} (Q1-Q4) This section collected developers' professional background to ensure relevance and interpretability of their responses. Developers were asked about their primary domain (e.g., UAVs, robotics, LLMs) (Q1), responsibility (hardware/software/algorithm development, testing, research, etc.) (Q2), years of experience (Q3), and geographical location (Q4).

\textbf{Practices and Challenges.}  (Q5-Q6) We explored whether the developers had ongoing projects combining LLMs and UAVs (Q5), and if not, what challenges (e.g., cost, technical limitations, regulation, unclear business models) had prevented integration (Q6).

\textbf{Application Maturity and Integration Approach.}  (Q7-Q10) Developers were asked to rate the maturity of various LLM application scenarios in UAVs, such as \textit{Natural Language Command Parsing}, \textit{Multimodal Information Parsing}, \textit{Single-UAV Task Reasoning and Planning}, and \textit{Multi-UAV Task Reasoning and Planning}, as well as other related tasks (Q7). Additional open-ended questions inquired about other potential applications desired by developers (Q8), the implementation approaches of UAV-LLM integration(e.g., \textit{External}, \textit{Hybrid}, and \textit{Internal}) (Q9), and the LLM's capabilities desired for UAV tasks (Q10).

\textbf{Academia-Industry Comparison.}  (Q11-Q13) To understand the alignment between research and practice, we asked whether developers referred to research work in their projects, and collected their views on the differences between academic and industrial efforts (Q11). Specific factors such as development lifecycles, technical requirements (Q12), and integration strategies were covered. 
Furthermore, we investigated the perceived underlying reasons for these differences, such as real-world requirements for technical maturity, performance, safety, and cost (Q13).

\subsubsection{Developer Recruitment.}
Our target practitioners for the survey were UAV developers. Our goal was to distribute the survey to GitHub developers who had contributed to UAV-related projects. First, we searched for UAV-related repositories on GitHub using keywords such as ``UAV''. Then, we used the GitHub Search API~\cite{github_api_docs} to collect the email addresses and names of contributors to the selected repositories. In total, we obtained 802 developer email addresses and sent out invitations to participate in the survey. Additionally, we shared the survey with several industry collaborators from well-known companies, including China Mobile Communications Corporation.

\subsubsection{Data Analysis.}
We received \textit{55} valid responses from \textit{15} countries (6.85\% response rate). After excluding three respondents without relevant experience, we analyzed 52 developers with an average of \textit{5.9} years of development experience. Their responsibilities included hardware development (\textit{12}, \textit{23.01\%}), software development (\textit{37}, \textit{71.15\%}), algorithm development (\textit{25}, \textit{48.08\%}), testing (\textit{9}, \textit{17.31\%}), and research (\textit{13}, \textit{25\%}).
Note that a developer may concurrently hold multiple roles, such as developing and testing software.

\vspace{-1em}
\subsection{Results}

\subsubsection{Practices and Challenges in UAV-LLM integration}

\textbf{Real-world Practices.}
In the survey, 40.4\% of developers have initiated projects integrating LLMs with UAVs. Among them, 19.05\% have fully deployed the technique, while 80.95\% remain in the proof-of-concept phase. This suggests that while integration is exploratory, initial applications have been successfully implemented. Conversely, 59.6\% of developers have not yet initiated such projects. We further inquired about the specific reasons.

\textbf{Challenges Perceived by Developers.}
The survey results identify the main obstacles that hinder the integration of UAVs and LLMs: existing techniques already meet the needs (54.8\%), existing LLMs cannot meet performance requirements (51.6\%), LLMs' security risks (45.2\%),  lack of relevant knowledge and technique(29.0\%), high computational costs (25.8\%), and lack of training data or fine-tuning methods (22.6\%). Furthermore, a few developers have noted that \textit{``LLMs are not being used on board in safety-critical applications''} and \textit{``LLMs should not be on a critical system due to their `black box' nature.''}

These findings reflect challenges to adoption. The high consensus on technical sufficiency suggests that traditional UAV solutions remain mature and competitive for current tasks, reducing the urgency for LLM-based alternatives. However, for advanced applications, a critical performance-safety gap persists. Developers are constrained by the limited real-time computational power of LLMs and their lack of interpretability in safety-critical operations. Furthermore, the convergence of high hardware costs, lack of domain-specific training data, and expertise shortages collectively diminishes the cost-effectiveness and feasibility of transitioning LLM research into real-world UAV projects.

\begin{center}
\begin{tcolorbox}[colback=gray!10,
                  colframe=black,
                  width=\columnwidth,
                  arc=1mm, auto outer arc,
                  boxrule=0.5pt,
                 ]
\noindent
\emph{\textbf{Finding 7:}  
Currently, 40.4\% of developers have attempted to integrate LLMs into UAVs.
The main reasons for not attempting are ``Existing technique has met the project requirements'' (54.8\%), ``Existing LLMs cannot meet performance requirements'' (51.6\%), and ``Security risks of LLMs'' (45.2\%). }
\end{tcolorbox}
\end{center}

\begin{center}
\begin{tcolorbox}[colback=gray!10,
                  colframe=black,
                  width=\columnwidth,
                  arc=1mm, auto outer arc,
                  boxrule=0.5pt,
                 ]
\noindent
\emph{\textbf{Finding 8:} A few developers have raised concerns about the application of LLMs in critical systems such as UAVs, indicating that integration is still in the early stages. While some developers have started applying them, demand in real-world UAV projects remains limited. To advance this field, it is necessary to enhance the performance, scalability, and reliability of LLMs in critical tasks.
 }
\end{tcolorbox}
\end{center}

\vspace{-1em}
\begin{table}[htbp]
    \centering
    \footnotesize 
    \renewcommand{\arraystretch}{1.2}
    \setlength{\tabcolsep}{6pt}
    
    \caption{Developers' perceptions on task maturity}
    \label{tab:task_maturity}

    \begin{tabular}{l c c} 
        \toprule
        \textbf{Reasons} & \textbf{Average} & \textbf{Score $\ge$ 3} \\ 
        \midrule
        Natural Language Command Parsing           & 2.89 & 57.7\% \\
        Multimodal Information Parsing            & 2.88 & 51.9\% \\
        Single-UAV Task Reasoning and Planning    & 2.77 & 48.1\% \\
        Multi-UAV Task Reasoning and Planning     & 2.40 & 38.5\% \\
        RAG-based Decision Support                & 2.88 & 55.8\% \\
        Control Logic and Command Generation      & 2.51 & 40.4\% \\
        Control Logic and Command Optimization    & 2.59 & 44.2\% \\
        Information Summary and Report Generation & 2.96 & 59.6\% \\
        LLM-based Chatbot                         & 2.96 & 61.5\% \\
        \bottomrule
    \end{tabular}
\end{table}

    


\textbf{Application Maturity Perceived by Developers.}
To assess developers' current perceptions of large models in UAVs, we invited developers to rate the technical maturity of the various tasks. We used the Likert scale~\cite{joshi2015likert} (1: completely immature; 3: neutral; 5: very mature). The results, as shown in ~\ref{tab:task_maturity}, indicate that the average score for the technical maturity of all tasks is below 3, indicating that while LLMs hold significant application potential in these areas, they also face numerous challenges. Among these, developers considered LLMs to have relatively higher technical maturity in \textit{Natural Language Command Parsing} (score 2.89) and \textit{Information Summarization and Report Generation} (score 2.96). These tasks involve basic processing of language and data, and existing LLMs can provide effective support in these areas, resulting in a relatively higher technical maturity. However, for more complex tasks, such as \textit{Multi-UAV Task Reasoning and Planning} (score 2.40) and \textit{Control Logic and Command Generation} (score 2.51), the technical maturity was lower. This suggests that, while multi-UAV coordination and autonomous control theoretically have significant potential, there are still many technical obstacles in effectively applying LLMs to these complex tasks in real-world UAV projects. These challenges need further improvement in areas such as accuracy, real-time performance, and collaborative computing capabilities. 

\begin{center}
\begin{tcolorbox}[colback=gray!10,
                  colframe=black,
                  width=\columnwidth,
                  arc=1mm, auto outer arc,
                  boxrule=0.5pt,
                 ]
\noindent
\emph{\textbf{Finding 9:} \textit{Information Summarization and Report Generation} (score 2.96) and \textit{Natural Language Command Parsing} (score 2.89) are perceived as the most mature tasks for LLM integration. In contrast, Multi-UAV Task Reasoning and Planning (score 2.40) and Control Logic and Command Generation (score 2.51) scored lower, suggesting that these tasks require higher accuracy, real-time performance, and computational resources, and current LLMs have yet to meet the decision support and real-time response requirements of existing UAVs. 
This leads to the low usage ratio of LLMs in UAV control tasks mentioned in Finding 6.}
\end{tcolorbox}
\end{center}

\textbf{Common UAV-LLM Integration Approaches.}
To understand the technical details of UAV-LLM integration, we asked developers to select from three integration methods. Survey results show that \textit{Hybrid-Onboard Assistance} is the most prevalent, chosen by 82.7\% of developers. In this approach, LLMs serve as supporting tools for task planning, path reasoning, and decision support rather than directly intervening in flight control.
In comparison, \textit{External-Offboard Invocation} (i.e., deploying large models on cloud or ground devices to provide decision support via external interfaces) and \textit{Internal-Onboard Control} (i.e., embedding large models into the UAV's control system to generate flight control commands directly) were selected by 67.3\% and 26.9\% of developers, respectively. This suggests that LLM integration is in its early stages, with developers believe that LLMs are better suited to supporting tasks rather than directly participating in flight control.

The restriction of LLMs to low-risk auxiliary tasks stems from concerns regarding their stability, interpretability, and real-time performance. 
Some developers in the survey pointed out, \textit{``I do not think LLMs are needed for tasks inboard the uav.''} and 
\textit{``AI can have a place on board for image processing to add sensors. But AI should never be put in direct control of a uav. It is an unnecessary, lazy, and non-robust approach to vehicle control.''} This clearly reflects the developers' cautious attitude towards the use of LLMs in critical UAV control tasks.

\textbf{The Need of UAV-LLM integration.}
Additionally, we asked developers about the level of need for LLMs in different UAV tasks. The results show that developers have a higher demand for \textit{Information Input} (score 2.42) and \textit{Task Planning} (score 2.53). This indicates that LLMs have significant application potential in processing user command translation, perception data, and Task Planning. These tasks require high data processing, reasoning, and planning capabilities. In contrast, the demand for \textit{Control Execution} (score 2.05) is lower, suggesting that traditional control algorithms remain dominant in real-time systems where existing techniques meet current needs or current LLMs cannot be fully relied upon.

\begin{center}
\begin{tcolorbox}[colback=gray!10,
                  colframe=black,
                  width=\columnwidth,
                  arc=1mm, auto outer arc,
                  boxrule=0.5pt,
                 ]
\noindent
\emph{\textbf{Finding 10:} Regarding task demands, Information Input (score 2.42) and Task Planning (score 2.53) received higher scores, and 82.7\% of developers believe that Onboard Assistance is the primary way to integrate large language models with UAVs. 
This indicates that, at present, LLMs in the UAV domain are more focused on supportive tasks rather than directly intervening in the flight control system.
This is consistent with the low maturity of tasks mentioned in Finding 9 and the cautious attitude of developers towards core control tasks in Finding 6.}
\end{tcolorbox}
\end{center}

\subsubsection{Researchers-Developers Comparison}

\textbf{Developer's Perception.}
To understand the practical application of academic research in industrial settings, we asked developers whether they refer to academic research in their actual work and included an open-ended question asking for the reasons. Results show that 69.2\% of developers refer to academic research to gain cutting-edge knowledge, verify information accuracy, accelerate development, and learn advanced algorithms. For example, one developer mentioned, ``\textit{I am implementing algorithms for paths' planning, etc. from the academic publications, I also use them during brainstorming of new ideas and to learn state-of-the-art.}''

\begin{center}
\begin{tcolorbox}[colback=gray!10,
                  colframe=black,
                  width=\columnwidth,
                  arc=1mm, auto outer arc,
                  boxrule=0.5pt,
                 ]
\noindent
\emph{\textbf{Finding 11:} 69.2\% of developers stated that they would reference academic research results in their actual work to stay updated on the latest improvements, suggesting that academic research plays an important role in driving UAV-LLM practices.}
\end{tcolorbox}
\end{center}

We also asked developers for their views on the differences between academia and industry. 
51.9\% of developers perceive industry as progressing faster than academia, particularly in practical application and implementation. 
In contrast, 36.5\% of developers believe that academia is progressing faster than industry. Academia's breakthroughs in basic theoretical research and innovative methods provide new ideas and directions for the industry. This divergence often stems from the need for misalignment between academic studies and practical industrial applications. As one developer noted, ``Industry tends not to be good at starting low trl development. In my experience, academia often develops things with little to no industrial application.'' This reflects the divergent requirements for Technology Readiness Levels (TRLs) between industry and academia, resulting in a disconnect between the two.

Additionally, 38.5\% of developers believe that there are differences between academia and industry in the integration methods. 67.3\% of developers think that academia and industry focus on different directions. Some developers pointed out: \textit{``Research - one topic, but deep, etc. industry - a lot of topics skimmed, practical usage with `black-box-magic'''}. This illustrates that researchers typically focus on deeply solving specific problems and exploring unknowns and theoretical innovations, while developers are more concerned with practical applications, focusing on how to apply quickly, efficiently, and reliably.

\begin{table*}[htbp]
    \centering
    \small 
    \renewcommand{\arraystretch}{1.25} 
    
    \caption{Developers' perceptions on the reasons for differences}
    \label{tab:dif_reasons}

    \newcolumntype{L}{>{\raggedright\arraybackslash}X}

    \begin{tabularx}{\textwidth}{l L c}
        \toprule
        \textbf{Challenge} & \textbf{Definition} & \textbf{Proportion} \\
        \midrule

        \multirow[t]{3}{*}{\textbf{Technical Maturity}} 
        & Actual projects require mature and stable technique (75.0\%) & \multirow[t]{3}{*}{84.6\%} \\
        & Difficulty in actual project development and integration (55.8\%) & \\
        & Immaturity of lightweight large models (40.4\%) & \\
        \cmidrule{1-3} 

        \multirow[t]{2}{*}{\textbf{Performance}} 
        & High computing resource requirements (32.7\%) & \multirow[t]{2}{*}{48.1\%} \\
        & Conflict between latency and real-time requirements (34.6\%) & \\
        \cmidrule{1-3}

        \multirow[t]{2}{*}{\textbf{Safety}} 
        & Model robustness and reliability challenges (30.7\%) & \multirow[t]{2}{*}{42.3\%} \\
        & Data privacy and security issues (19.2\%) & \\
        \cmidrule{1-3}

        \multirow[t]{2}{*}{\textbf{Cost}} 
        & High computing power/hardware costs (46.2\%) & \multirow[t]{2}{*}{44.2\%} \\
        & Commercial input-output ratio factors (28.8\%) & \\
        \cmidrule{1-3}

        \multirow[t]{3}{*}{\textbf{Other}} 
        & Academic research is out of touch with industrial needs (36.5\%) & \multirow[t]{3}{*}{57.7\%} \\
        & Regulatory and policy uncertainty (21.2\%) & \\
        & Lack of large model training data (25.0\%) & \\

        \bottomrule
    \end{tabularx}
\end{table*}

\textbf{Solutions Desired by Developers.}
We asked developers to identify the reasons and solutions for these differences. The responses can be categorized into the five challenges listed in ~\ref{tab:dif_reasons}. Notably, 84.6\% of developers identified ``Technical Maturity'' as the main challenge, reflecting the industry's emphasis on reliable solutions over theoretical innovation. Specifically, developers pointed to the need to address issues such as ``Actual projects require mature and stable technique'' (75.0\%), ``Difficulty in actual project development and integration'' (55.8\%), and ``Immaturity of lightweight large models'' (40.4\%). These findings reflect the challenges academic research faces in translating results into real-world applications, as industry often encounters complex conditions during project development and integration.

Developers also noted that industry faces challenges related to ``Performance'' (48.1\%), ``Safety'' (42.3\%), ``Cost'' (44.2\%), and ``Other'' (57.7\%). Corresponding issues include ``High computational resource requirements'' (46.2\%), ``Conflicts between latency and real-time requirements'' (34.6\%), ``High computation/hardware costs'' (32.7\%), "Model robustness and reliability challenges" (30.7\%), and ``Commercial return on investment'' (28.8\%). One developer emphasized: \textit{``Industry wants practical results fast, especially smaller startups, etc. Academia has smaller groups to work with, less money, etc.''} This highlights the urgent demand for rapid commercialization in industry, which makes the transfer of academic research outcomes into practical industrial applications more challenging.
\vspace{-1em}

\begin{center}
\begin{tcolorbox}[colback=gray!10,
                  colframe=black,
                  width=\columnwidth,
                  arc=1mm, auto outer arc,
                  boxrule=0.5pt,
                 ]
\noindent
\emph{\textbf{Finding 12:} Developers believe that the differences between academia and industry primarily lie in their focus areas (67.3\%) and technical requirements (63.5\%). The greatest challenge in addressing these differences is technical maturity (84.6\%).
While academic work drives innovation, real-world integration is more complex and requires mature, stable, and practically feasible solutions.}
\end{tcolorbox}
\end{center}

\section{Discussion}
\label{sec:ds}
This section explains the significance of this study for researchers and developers, summarizes the three main challenges of UAV--LLM applications, and presents the validity threats.

\vspace{-1em}
\subsection{Implications}
\noindent \textbf{For researchers.}
Our study reveals the significant differences between researchers' and developers' practices in integrating UAVs with LLMs. Researchers focus on theoretical modeling and the exploration of new approaches (e.g., 52.7\% using \textit{External-Offboard Invocation} modes), while developers are more concerned with practical applications and system integration (e.g., 50\% using \textit{Hybrid-Onboard Assistive} modes). To bridge this gap, researchers should align with real-world needs regarding real-time performance, robustness, and safety. Priority should be given to developing lightweight models for resource-constrained devices~\cite{edgeLLM2025survey,xu2024ondevicelanguagemodelscomprehensive}, particularly for complex tasks such as \textit{Multi-UAV Task Reasoning and Planning} and \textit{Control Logic and Command Generation}. Additionally, researchers should enhance studies on Explainable AI (XAI)~\cite{Black-Box} and trust mechanisms to address industry developers' concerns over data privacy, model transparency, and security in critical tasks.

\noindent \textbf{For developers.}
This study clarifies the tasks that current LLMs can reliably support, such as task planning, user interaction, and automatic report generation. These tasks show stronger performance and lower implementation risks. The comparative analysis also indicates that many complex academic solutions are not yet ready for real-world use because of issues with latency, cost, and reliability. The results further point out important challenges in real-time processing and computational efficiency, giving developers a more realistic view of short-term priorities.
To this end, developers can move forward in two directions. First, they can apply LLMs in tasks that are already feasible, such as building more natural user interfaces, generating reports automatically, and supporting mission planning. These areas involve lower risks and can deliver practical value relatively quickly. Second, for tasks that are not yet suitable for deployment, developers can focus on researching and testing improvements. This includes integrating lighter models~\cite{xu2024ondevicelanguagemodelscomprehensive}, developing smaller or domain-specific models, or combining LLMs with traditional UAV control systems~\cite{brunke2022safe}. In this way, developers can benefit from the current capabilities of LLMs while also preparing step by step for more advanced applications in the future.

\vspace{-1em}
\subsection{Future Directions for UAV-LLM Applications}

\noindent \textbf{(1) Bridging the Academia-Industry Gap in UAV-LLM Integration:}
There is a mismatch between academic research and industrial solutions, with 67.3\% of developers believing the two fields focus on different directions and 63.5\% noting differing technical requirements. Researchers often prioritize theoretical problems, frequently overlooking industrial needs for real-time performance, stability, and system integration (75\%). Conversely, developers focus on commercialization and rapid deployment, slowing the transfer of academic outcomes. To accelerate technology transfer, future research should align with practical needs. Researchers can collaborate with industry to obtain real-world data and feedback, ensuring new technologies are effectively deployed while avoiding unnecessary obstacles during transfer.

\noindent \textbf{(2) Improving the Safety and Reliability of LLMs to Enhance Developer Trust:}
Deploying LLMs on UAVs faces obstacles related to safety and reliability. Our survey shows that 45.2\% of developers highlight the security risks posed by large models, while 82.7\% prefer using LLMs in supportive roles. Their interest in using LLMs for direct control execution is low (score: 2.05). Some developers further emphasized that ``LLMs should not be on a critical system due to their `black box' nature'' and that ``AI should never be put in direct control of a UAV.'' These concerns stem from a lack of interpretability and transparency, leading to uncertainty and potential system failures. Furthermore, they reflect a reliance on traditional control algorithms rather than LLMs for decision-making, since current LLMs are viewed as unable to meet performance requirements (51.6\%) and their capability in control logic generation remains immature (lowest score: 2.51). Future research should enhance model interpretability and transparency to increase developer trust. In addition, it should also develop efficient, lightweight models with lower computational demands to meet the real-time processing and computational requirements of complex tasks on UAV platforms.

\vspace{-1em}
\subsection{Threats To Validity}

\subsubsection{Internal Validity}

During the process of literature screening and classification, researchers' subjective judgments may influence the inclusion criteria and labeling of studies, potentially introducing bias into the analysis. Two authors—each with over two years of research experience in the UAV field—independently conducted literature selection and labeling. All disagreements were resolved through thorough discussion and consensus, ensuring the validity and consistency of the analysis.
To reduce the risk of unprofessional or dishonest responses, the survey was conducted anonymously, explicitly informing participants that no personally identifiable information would be collected~\cite{ong2000impact}. The questionnaire primarily consisted of multiple-choice items with an ``I don't understand this question'' option, and developers were free to skip ambiguous items. This design aimed to enhance data quality while protecting participant privacy.
To collect more responses and overcome language barriers, we translated the questionnaire into Chinese and distributed it via platforms accessible in mainland China. While we made every effort to ensure consistency between the English and Chinese versions, minor discrepancies may have arisen during translation, which could affect developers' understanding. To minimize this risk, two Chinese authors with strong English proficiency collaboratively reviewed and repeatedly refined the translations to ensure their accuracy and consistency.

\subsubsection{External Validity}

This study collected survey responses from active GitHub developers. However, it is possible that certain groups of developers, such as those working on proprietary systems or those less active in the open-source community, were not adequately covered, and their perspectives may differ from those represented in our sample. Nevertheless, our sample exhibits substantial diversity in UAV-related experience, professional roles, and levels of project involvement, enhancing confidence in the generalizability of the findings.
Additionally, as UAV platforms and LLMs advance rapidly, future systems may integrate new tasks or optimized architectures, potentially making some specific findings outdated over time. This is an inherent limitation of empirical research in rapidly developing technological domains. Despite this, our proposed methodology and task framework remain broadly applicable, providing a timely and valuable reference for researchers and developers seeking to advance UAV-LLM integration.

\section{Related Work}
\label{sec:rw}
\subsection{UAV-LLM Integration}

A growing body of research has investigated concrete systems and technical frameworks for integrating LLMs with UAVs. Chen ~\textit{et al.}\cite{chen2023typefly} developed TypeFly, the first real-world UAV platform integrated with an LLM to support natural language-driven flight control. Their system demonstrated the feasibility of mapping natural language to motion commands but also revealed limitations in handling semantic ambiguity and adapting to dynamic environments. Phadke ~\textit{et al.}\cite{phadke2024integrating} proposed a modular UAV control framework based on LLMs in simulation, coordinating navigation, obstacle avoidance, and target recognition modules. While the modular design improved stability, the absence of a unified semantic memory layer restricted decision efficiency. Zhao and Lin~\cite{zhao2025general} introduced the concept of a Universal Aerial Agent with the ``Prompt-as-Policy'' approach, enabling flexible encoding of task intents, though their work remains at a conceptual stage without large-scale validation. More recently, Koubaa~\cite{koubaa2025next} introduced UAV-GPT, which translates high-level semantic instructions into executable UAV tasks, while Ren ~\textit{et al.}~\cite{ren2024grounded} integrated foundation models with UAV perception for semantic annotation and visual reasoning.

In parallel, several works examine UAV software quality and system reliability, which are important as intelligent control strategies emerge. Di Sorbo ~\textit{et al.} \cite{di2023automated} analyzed safety concerns in UAV platforms, highlighting issues from fragile system design. For runtime monitoring, Shar ~\textit{et al.}\cite{shar2022dronlomaly} introduced DronLomaly for anomaly detection through log analysis. Wang ~\textit{et al.}\cite{wang2025routhsearch} proposed RouthSearch to infer PID parameter specifications for flight-control programs, helping improve the robustness of low-level control. Furthermore, Wang ~\textit{et al.}\cite{wang2024exploratory,wang2021exploratory}conducted exploratory studies on UAV log anomalies and autopilot software bugs, offering empirical insights into common reliability problems in existing UAV systems.

These technical explorations highlight the potential of LLMs to enhance the autonomy, perception, and interaction of UAVs. However, most existing systems are still early prototypes without large-scale validation. The software-engineering studies discussed above also indicate that safety, anomaly detection, and reliable control are necessary foundations for practical UAV-LLM integration. Compared to these works, our study provides a more comprehensive perspective: we categorize tasks across the UAV workflow, compare differences between academic and industrial practices, and identify key challenges reported by developers. This broader view helps to contextualize existing prototypes within a systematic landscape and points out where research and practice can be better aligned.

\vspace{-1em}
\subsection{Empirical Studies on UAVs}

A number of empirical studies have summarized the use of machine learning techniques in UAVs. For example, Osco ~\textit{et al.}\cite{osco2021review} reviewed the role of deep learning in UAV remote sensing, covering tasks such as image classification, object recognition, and environmental monitoring. Tang ~\textit{et al.}\cite{tang2023survey} provided a survey on object detection for UAVs based on deep learning, emphasizing algorithmic advances and benchmark datasets. AlMahamid and Grolinger~\cite{almahamid2022autonomous} systematically reviewed reinforcement learning approaches for autonomous UAV navigation, highlighting progress and challenges in dynamic environments. Bai ~\textit{et al.}~\cite{bai2023toward} focused on reinforcement learning--based methods for building autonomous multi-UAV wireless networks, summarizing techniques for coordination and communication. These surveys clearly show the important role of machine learning techniques in UAV research, but none of them address the emerging opportunities and challenges brought by large language models. In other words, although UAV-DL/ML integration has been extensively reviewed, the UAV--LLM integration landscape remains largely unexplored.

There has been one recent attempt to discuss UAV-LLM integration. Javaid ~\textit{et al.}~\cite{javaid2024large} reviewed possible applications of LLMs in UAVs, focusing on natural language command control, reasoning-assisted decision making, and human-computer interaction. However, their work mainly focuses on conceptual and qualitative descriptions, without connecting academic, industry, and developers' perspectives. In contrast, our study constructs the first taxonomy of UAV-LLM tasks, analyzes 997 academic papers and 1,509 open-source projects, and further validates our findings with survey responses from 52 developers. This large-scale effort not only summarizes existing research but also quantifies the gap between academia and industry, offering practical guidance for real-world deployment.

\section{Conclusion}
\label{sec:conclusion}
In this paper, we conduct an empirical study of LLM applications in UAVs to examine the current state of research and development. First, by analyzing 74 papers and 56 open-source GitHub projects, we summarize nine categories of LLM-supported tasks across four UAV workflows. Second, we quantify the distribution of these tasks in researchers' and developers' practices and perform a comparative analysis. The results reveal gaps between researchers and developers in research focus, technical requirements, and integration modes of LLMs and UAVs. We then administer an online survey and obtain 52 valid responses. We find that 59.6\% of developers have not yet attempted to use LLMs in real-world UAV projects. Their feedback indicates that current LLMs fall short of meeting the requirements of UAVs for accuracy, real-time performance, and reliability, and that the lack of technical maturity and practical feasibility in academic research further increases the complexity of engineering deployment. Finally, based on our findings, we highlight future challenges for the application of LLMs in UAVs and offer recommendations for researchers and developers.

\bgroup
\def\baselinestretch{0.93}
\bibliographystyle{IEEEtran}
\bibliography{ref}
\egroup

\vfill
\end{document}